# A Digital SRAM-Based Compute-In-Memory Macro for Weight-Stationary Dynamic Matrix Multiplication in Transformer Attention Score Computation

Jianyi Yu, *Student Member*, *IEEE*, Tengxiao Wang, Yuxuan Wang, Xiang Fu, *Student Member*, *IEEE*, Fei Qiao, Ying Wang, Rui Yuan, Liyuan Liu, *Member, IEEE*, Cong Shi*, *Member*, *IEEE*

***Abstract*—Compute-in-memory (CIM) techniques are widely employed in energy-efficient artificial intelligent (AI) processors. They alleviate power and latency bottlenecks caused by extensive data movements between compute and storage units. To extend these benefits to Transformer, this brief proposes a digital CIM macro to compute attention score. To eliminate dynamic matrix multiplication (MM), we reconstruct the computation as static MM using a combined QK-weight matrix, so that inputs can be directly fed to a single CIM macro to obtain the score results. However, this introduces a new challenge of 2-input static MM. The computation is further decomposed into four groups of bit-serial logical and addition operations. This allows 2-input to directly activate the word line via AND gate, thus realizing 2-input static MM with minimal overhead. A hierarchical zero-value bit skipping mechanism is introduced to prioritize skipping zero-value bits in the 2-input case. This mechanism effectively utilizes data sparsity of 2-input, significantly reducing redundant operations. Implemented in a 65-nm process, the 0.35 mm² macro delivers 42.27 GOPS at 1.24 mW, yielding 34.1 TOPS/W energy and 120.77 GOPS/mm² area efficiency. Compared to CPUs and GPUs, it achieves ~25× and ~13× higher efficiency, respectively. Against other Transformer-CIMs, it demonstrates at least 7× energy and 2× area efficiency gains, highlighting its strong potential for edge intelligence.**

***Index Terms*—Transformer, attention, weight-stationary, 2-input static MM, compute-in-memory, energy-efficient**

## I. INTRODUCTION

The advancement of the Transformer technology, driven by algorithmic innovations and hardware capabilities, has been a significant catalyst for progress in artificial intelligence. However, its massive computational demands and data movement create significant hardware performance bottlenecks, exacerbating the "memory wall" issue inherent in von Neumann architectures. In response, the research community is investigating compute-in-memory (CIM) techniques. This shift from processor-centric to data-centric computing integrates multiplication and accumulation (MAC) operations within the memory array, effectively merging processing and storage. This architecture aims to minimize data movement, delivering highly energy-efficient matrix-vector multiplication for deep neural networks [1], [2].

The digital CIM architecture amalgamates the respective strengths of CIM and conventional digital design. It achieves high efficiency by directly embedding computation within the memory fabric, while simultaneously leveraging its inherent digital logic to eliminate the non-idealities associated with analog circuits, thereby ensuring high computational accuracy [3]-[5]. However, the introduction of the attention mechanism in Transformer models presents new challenges for existing digital CIM accelerator designs.

*Challenge 1*: As shown in Fig. 1, The query (**Q**) and key (**K**) matrices are dynamically generated in the attention mechanism. Because of this dynamic generation, the computation of score matrix ($\mathbf{S}=\mathbf{Q}\times\mathbf{K}^T$) becomes dynamic matrix multiplication (MM). The emergence of dynamic MM directly undermines the advantage of traditional CIM architectures that rely on a weight-stationary method for static MM, thereby necessitating frequent updates to the memory data in the CIM. Although existing Transformer CIM architectures [6]-[10] reduce redundant off-chip accesses through pipelined or ring-shaped structures, they do not completely address the overhead associated with updating the CIM's memory data. Moreover, the computation of **S** requires multiple CIMs to work together.

*Challenge 2*: While the $\mathbf{QK}^T$ pre-calculation effectively addresses the challenge posed by *Challenge 1*, it introduces new design challenge. As illustrated in Fig. 1, traditional CIMs can only implement the **XW** computation form [1], [11], which involves a single input **X**. However, after the $\mathbf{QK}^T$ pre-calculation, the computation of **S** becomes closely related to the **XWY** computation form, which involves 2-input: **X** and **Y**. This shift leads to a significant increase in the demand for memory access and data transfer, consequently resulting in higher latency and energy consumption. Furthermore, handling computations with multiple inputs further increases the design complexity.

*Challenge 3*: Zero-value bit skip can leverage the sparsity of input data to reduce redundant computations. However, in Challenge 2, while the approach of using 2-input driven word line activation aims to minimize the complexity, the logical operations between the 2-input result in an increase in input sparsity. This hinders the traditional 1-input zero-value bit skip from fully leveraging input sparsity. For 2-input cases, the skip

This work was funded in part by the Major Research Plan of the National Natural Science Foundation of China under Grant No. 92464103, and in part by the Key Program of the National Natural Science Foundation of China under Grant No. 6233400.

Jianyi Yu, Tengxiao Wang, Yuxuan Wang, Xiang Fu and Cong Shi and are with the School of Microelectronics and Communication Engineering, Chongqing University, Chongqing 400044, China. (Corresponding author: Cong Shi, e-mail: shicong@cqu.edu.cn)

Fei Qiao is with the Department of Electronic Engineering, Tsinghua University, Beijing 100084, China.

Ying Wang is with State Key Laboratory of Processors, Institute of Computing Technology, Chinese Academy of Sciences, Beijing 100190, China.

Rui Yuan is with the School of Artificial Intelligence, Southwest University, Chongqing, 402460, China.

Liyuan Liu is with the State Key Laboratory of Semiconductor Physics and Chip Technologies, Institute of Semiconductors, Chinese Academy of Sciences, Beijing 100083, China.

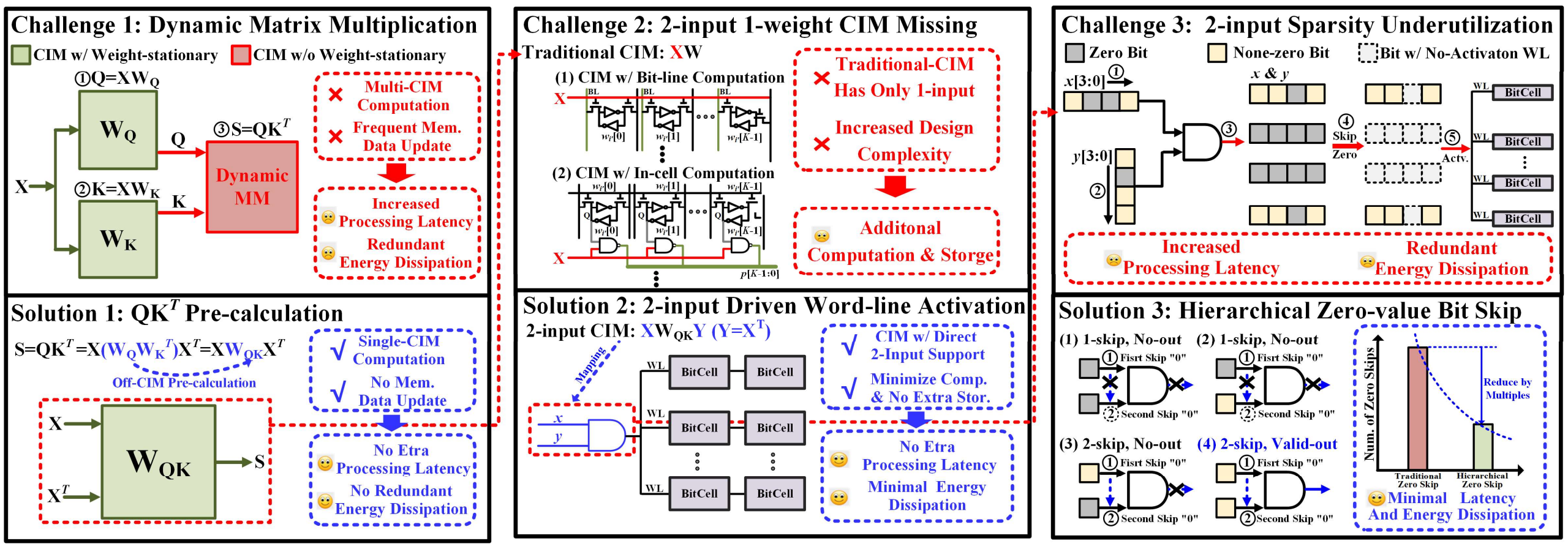

Fig. 1. The challenges for CIM-based accelerator designs in transformer's s attention score computation

operations on these inherently sparse inputs also incur unnecessary computational overhead.

In this brief, we propose a digital CIM macro for energy-efficiency computation in Transformer. The macro overcomes the above challenges with the following three features, as shown in Fig. 1.

1) For Challenge 1, we reformulate the attention score calculation based on a combined QK-weight matrix. By feeding the input data directly into a CIM macro employing a weight-stationary architecture, the result is obtained through static MM, which obviates the need for dynamic MM.

2) For Challenge 2, we decompose the attention score calculation into four groups of logical and addition operations. Thus, the 2-input can be directly processed through the AND-gate results, followed by activating the word line to perform the multiplication with the weights.

3) For Challenge 3, We adopt a hierarchical zero-value bit skip to improve computational efficiency and reduce redundant operations. This mechanism allows for the selective skipping of unnecessary computations at different levels, thereby better leveraging the sparsity of the input data.

Compared to other Transformer CIMs, experimental results demonstrate the CIM macro achieves at least 7× higher energy efficiency and 2× higher area efficiency at the same technology node, highlighting its advantages for edge intelligence.

## II. Algorithm Optimization

### A. Optimization via QK-Weight Pre-calculation

In the traditional Transformer, the calculation of $\mathbf{S}$ in the attention mechanism is presented [11] in Eq. (1):

$$\mathbf{S} = \mathbf{Q}\mathbf{K}^T = \mathbf{X}\mathbf{W}_\mathrm{Q}(\mathbf{X}\mathbf{W}_\mathrm{K})^T = \mathbf{X}\mathbf{W}_\mathrm{Q}\mathbf{W}_\mathrm{K}^T\mathbf{X}^T \tag{1}$$

where $\mathbf{S}\in\mathbb{R}^{N\times N}$ is the attention score matrix, $\mathbf{Q}\in\mathbb{R}^{N\times d}$ is the query matrix, $\mathbf{K}\in\mathbb{R}^{N\times d}$ is the key matrix, and $\mathbf{X}\in\mathbb{R}^{N\times d_\mathrm{model}}$ is the input matrix composed of $N$ tokens, each with a dimension of $d_\mathrm{model}$. $\mathbf{W}_\mathrm{Q}$ and $\mathbf{W}_\mathrm{K}\in\mathbb{R}^{D\times d}$ are the corresponding weight matrices.

In inference, $\mathbf{W}_\mathrm{Q}$ and $\mathbf{W}_\mathrm{K}$ are constant. As shown in Fig. 2, we pre-compute $\mathbf{W}_\mathrm{Q}$ and $\mathbf{W}_\mathrm{K}$, resulting in a new matrix $\mathbf{W}_\mathrm{QK}$:

$$\mathbf{W}_\mathrm{QK} = \mathbf{W}_\mathrm{Q}\mathbf{W}_\mathrm{K}^T \in \mathbb{R}^{D\times D} \tag{2}$$

Subsequently, we derive the following expression:

$$\mathbf{S} = \mathbf{X}\mathbf{W}_\mathrm{QK}\mathbf{X}^T \tag{3}$$

To compute each element of $\mathbf{S}$, $\mathbf{X}$ is flattened as follows:

$$\mathbf{X}=\begin{bmatrix}\mathbf{X}_1 & \mathbf{X}_2 & \cdots & \mathbf{X}_N\end{bmatrix}^T \tag{4}$$

Thus, $\mathbf{S}$ can be arranged as follows:

$$\mathbf{S} = \begin{bmatrix} \mathbf{X}_1\mathbf{W}_\mathrm{QK}\mathbf{X}_1^T & \mathbf{X}_1\mathbf{W}_\mathrm{QK}\mathbf{X}_2^T & \cdots & \mathbf{X}_1\mathbf{W}_\mathrm{QK}\mathbf{X}_N^T \\ \mathbf{X}_2\mathbf{W}_\mathrm{QK}\mathbf{X}_1^T & \mathbf{X}_2\mathbf{W}_\mathrm{QK}\mathbf{X}_2^T & \cdots & \mathbf{X}_2\mathbf{W}_\mathrm{QK}\mathbf{X}_N^T \\ \vdots & \vdots & \ddots & \vdots \\ \mathbf{X}_N\mathbf{W}_\mathrm{QK}\mathbf{X}_1^T & \mathbf{X}_N\mathbf{W}_\mathrm{QK}\mathbf{X}_2^T & \cdots & \mathbf{X}_N\mathbf{W}_\mathrm{QK}\mathbf{X}_N^T \end{bmatrix} \tag{5}$$

Here, $s_{ij}$ in the $i$-th row and $j$-th column of $\mathbf{S}$ is given by:

$$s_{ij} = \mathbf{X}_i\mathbf{W}_\mathrm{QK}\mathbf{X}_j^{\,T} \tag{6}$$

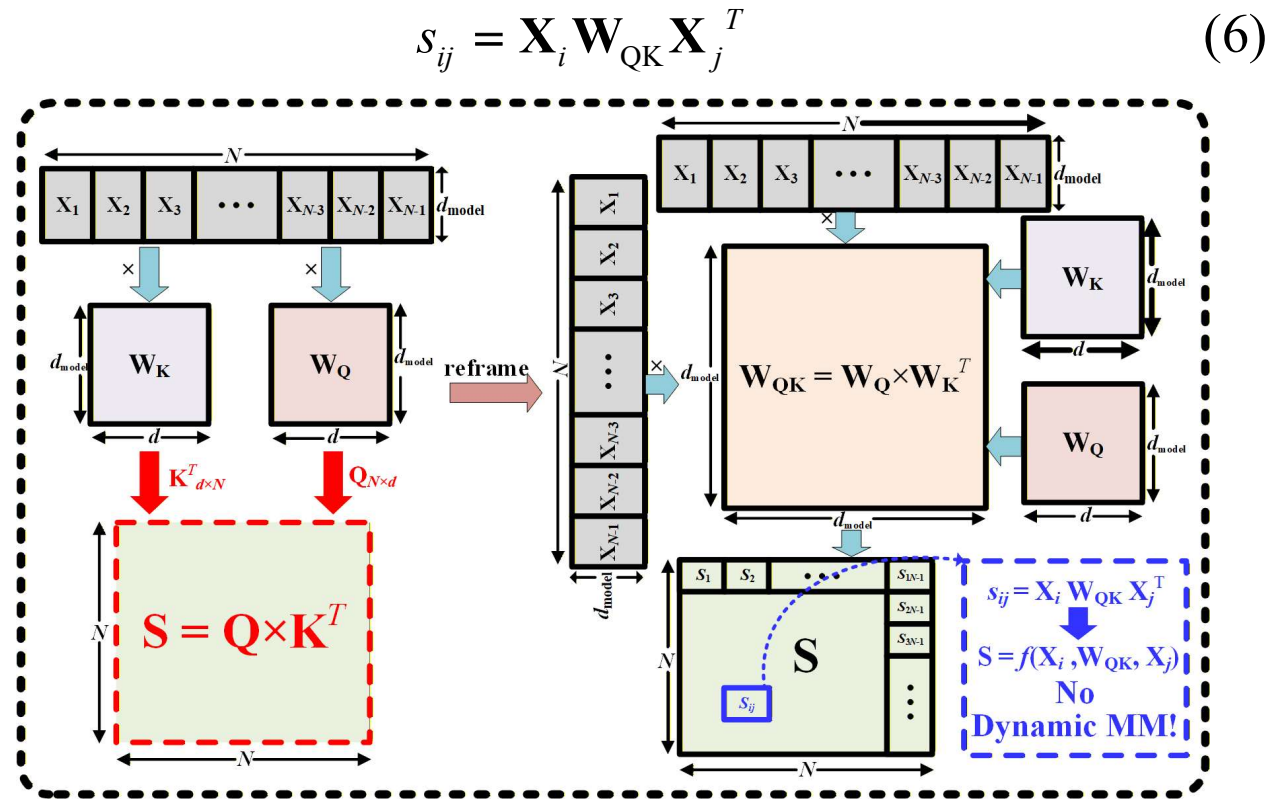

Fig. 2. The process of pre-computing $\mathbf{W}_\mathrm{K}$ and $\mathbf{W}_\mathrm{Q}$ Multiplication.

Therefore, each $s_{ij}$ of $\mathbf{S}$ can be directly computed from the pre-computed $\mathbf{W}_\mathrm{QK}$ and the corresponding input $\mathbf{X}_i$ and $\mathbf{X}_j^T$ through static MM, which avoids the need for dynamic MM.

### B. Bit-Slicing for Input Data

Eq. (6) is further expanded to derive the following computational expression for $s_{ij}$:

$$s_{ij}=\sum_{j'=1}^{D}\sum_{i'=1}^{D}x_{ii'}w_{\mathrm{QK},i'j'}x_{jj'} \tag{7}$$

where $x_{ii'}$ denotes the $i'$-th entry of the row vector $\mathbf{X}_i$, $x_{jj'}$ represents the $j'$-th entry of the column vector $\mathbf{X}_j$, and $w_{\mathrm{QK},i'j'}$ represents the element in the $i'$-th row and $j'$-th column of $\mathbf{W}_{\mathrm{QK}}$.

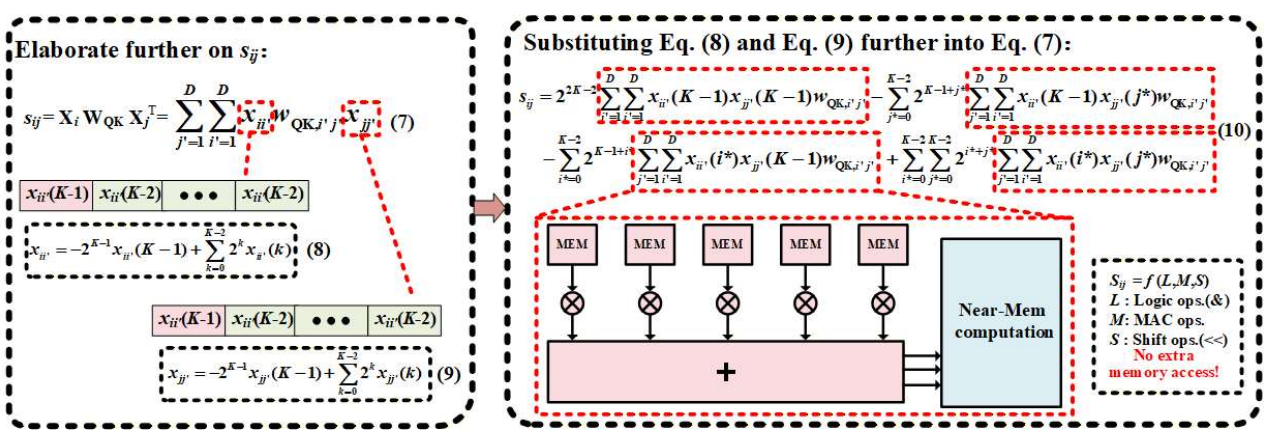

Fig. 3. The process of bit slicing divides input data into single bits

To implement CIM, we must express $x_{ii'}$ and $x_{jj'}$ in terms of their binary representations. Thus, we further expand:

$$x_{ii'}=-2^{K-1}x_{ii'}(K-1)+\sum_{k=0}^{K-2}2^{k}x_{ii'}(k) \tag{8}$$

$$x_{jj'}=-2^{K-1}x_{jj'}(K-1)+\sum_{k=0}^{K-2}2^{k}x_{jj'}(k) \tag{9}$$

where $K$ represents the bit width of $x_{ii'}$ and $x_{jj'}$, with the most significant bit serving as the sign bit. Substituting Eq. (8) and Eq. (9) into Eq. (7), we can obtain:

$$\begin{aligned}
s_{ij}&=\sum_{j'=1}^{D}\sum_{i'=1}^{D}\Big(-2^{K-1}x_{ii'}(K-1)+\sum_{i^*=0}^{K-2}2^{i^*}x_{ii'}(i^*)\Big)w_{\mathrm{QK},i'j'}\Big(-2^{K-1}x_{jj'}(K-1)+\sum_{j^*=0}^{K-2}2^{j^*}x_{jj'}(j^*)\Big)\\
&=2^{2K-2}\sum_{j'=1}^{D}\sum_{i'=1}^{D}x_{ii'}(K-1)x_{jj'}(K-1)w_{\mathrm{QK},i'j'}\\
&-\sum_{j^*=0}^{K-2}2^{K-1+j^*}\sum_{j'=1}^{D}\sum_{i'=1}^{D}x_{ii'}(K-1)x_{jj'}(j^*)w_{\mathrm{QK},i'j'}\\
&-\sum_{i^*=0}^{K-2}2^{K-1+i^*}\sum_{j'=1}^{D}\sum_{i'=1}^{D}x_{ii'}(i^*)x_{jj'}(K-1)w_{\mathrm{QK},i'j'}\\
&+\sum_{i^*=0}^{K-2}\sum_{j^*=0}^{K-2}2^{i^*+j^*}\sum_{j'=1}^{D}\sum_{i'=1}^{D}x_{ii'}(i^*)x_{jj'}(j^*)w_{\mathrm{QK},i'j'}
\end{aligned} \tag{10}$$

where $x_{ii'}(i^*)$ denotes the $i^*$-th bit of the $x_{ii'}$, and $x_{jj'}(j^*)$ denotes the $j^*$-th bit of the $x_{jj'}$, with $i^*, j^* = 0, 1, 2, \ldots, K-1$.

We decompose the calculation for 2-input into four logical, shift, and addition operations. This decomposition facilitates the progressive execution of the entire computation using bit-serial inputs, making it compatible with CIM design paradigm.

## III. Hardware Design

The proposed restructuring of attention score computation enables direct processing by a single CIM macro, avoiding dynamic matrix multiplication and extra macros. However, this poses a new architectural challenge, as traditional CIM architectures are designed for 1-input (X) form, while the new method necessitates 2-input (X, Y) form. As a core solution to this conflict, we introduce an innovative CIM macro capable of handling 2-input computations.

### A. 2-Input High-Performance and Scalable CIM Macro

As shown in Fig. 4, we have developed a 2-input highly efficient CIM macro, whose core architecture is composed of a CIM array and a near-memory computing module. The 2-input feature is realized through a sophisticated hardware mapping scheme: first, two input data streams ($x_{ii'}(i^*)$, $x_{jj'}(j^*)$) undergo a logical "AND" operation within the input buffer module. The result is then directly used as a data-driven signal to control the word lines of the SRAM array, thereby completing the parallel multiplication with the weight $w_{\mathrm{QK},i'j'}$ in a single step. The circuit we propose, which maps 2-input multiplication to "AND-gated driven word line activation", offers significant hardware design cost reductions compared to integrating 2-input logic within each SRAM cell. This scheme requires no extra transistors in the memory cell, thereby avoiding the area overhead and complexity arising from cell-level modifications, while preserving the inherent stability and manufacturability. CIM accumulates multiplication results over multiple cycles for each of the four polynomials. The near-memory module then computes $s_{ij}$ via intra-group shifting, accumulation, and arithmetic on the four result groups.

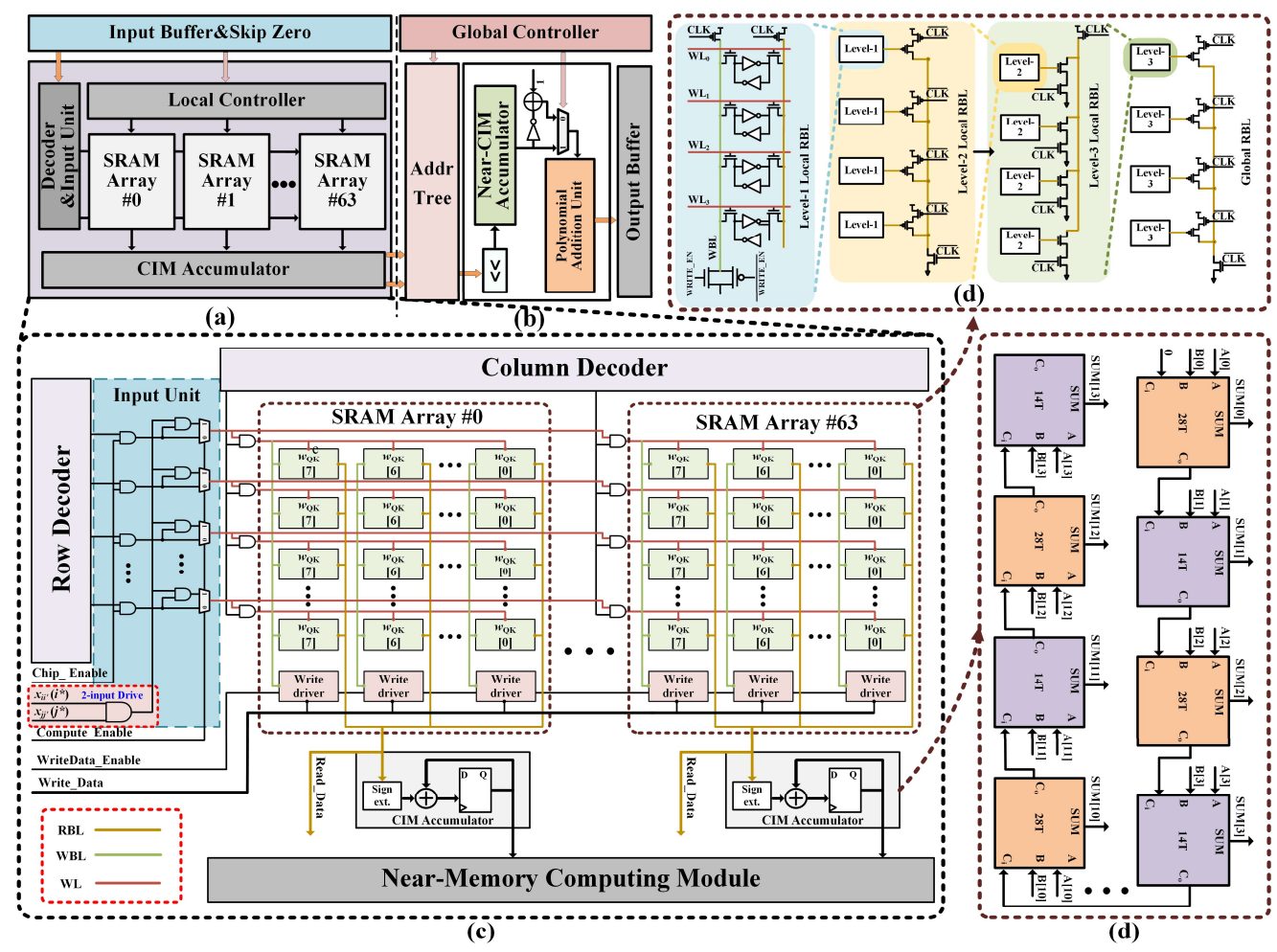

Fig. 4. (a) The proposed CIM macro's overall architecture. (b) The near-memory computing module. (c) The high-performance and scalable CIM bank. (d) The core computing circuit design.

To enhance the stability and energy efficiency of the CIM array, we design a read-write separated 6T-SRAM array to eliminate read disturbance, and enhanced the signal integrity in large-scale arrays using NP domino logic multi-level buffering. Concurrently, we implement a high-performance accumulator with an alternating 14T/28T full adder structure. This design both avoids the excessive delay of a 14T-only structure and reduces the area and power consumption inherent to a 28T-only structure. These optimizations work in concert to realize a highly stable and energy-efficient CIM macro.

### B. The Sparsity of Input Data and Zero-Skip Mechanism

A significant number of zero value bits appear in Transformer data, caused by a combination of padding (used to unify sequence lengths), the short-sequence nature of text, and low-frequency embedding. Additionally, activation functions such as GELU output many near-zero values, which further increases the data's bit sparsity. As illustrated in Fig. 5(a), to address this critical issue, Hierarchical zero-value bit skip provides a more efficient solution. By identifying and skipping zero-value bits at different data granularity levels, its collaborative decision-making mechanism ensures that a computation is triggered only when both inputs are non-zero at the corresponding level and position. This method, therefore, fundamentally circumvents invalid computations, thereby not only reducing computational waste and memory access overhead but also achieving significant reductions in overall power consumption and computational latency.

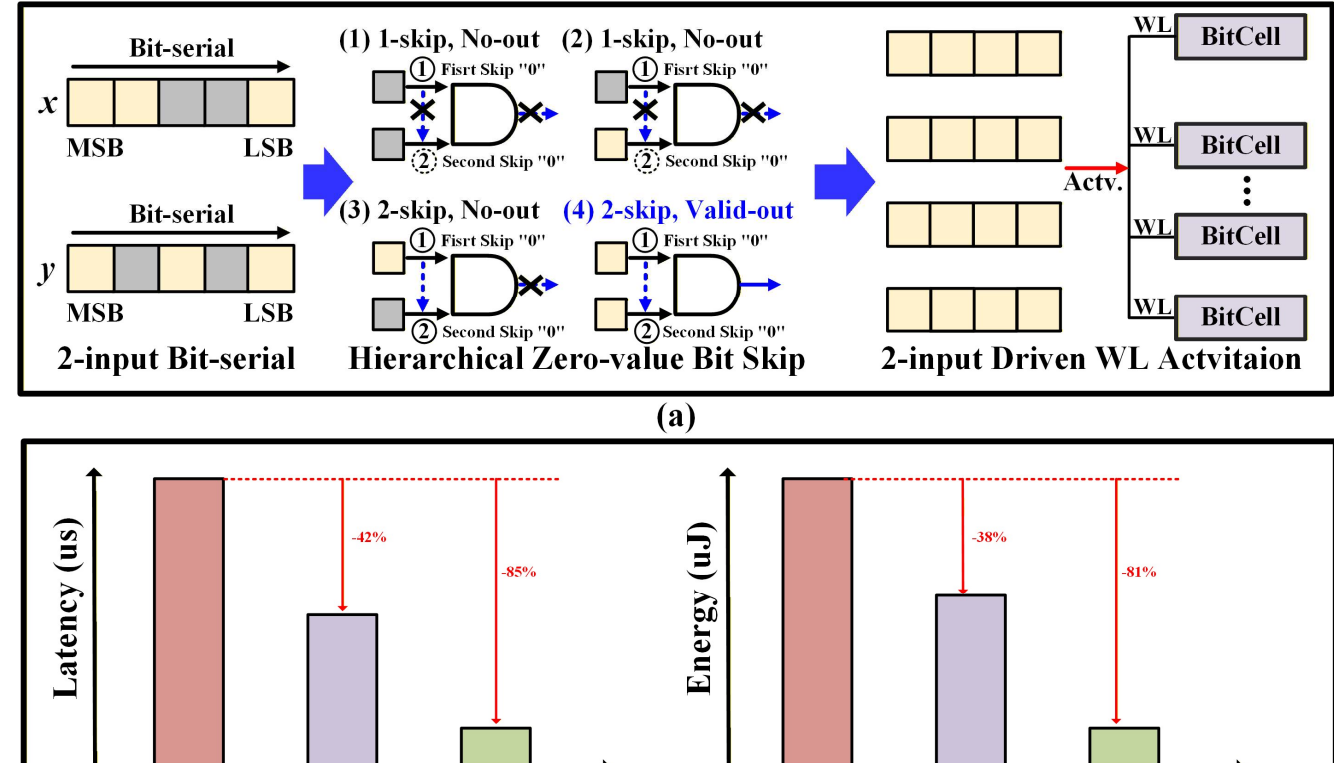

Fig. 5. (a) The hierarchical zero-value skip. (b) Evaluation for the hierarchical zero-value bit skip architecture on latency and energy.

To validate the mechanism, we used a hybrid simulation strategy: a Verilog behavioral model tallied total operations and computation cycles, while post-layout simulation of the CIM macro provided per-operation energy and latency benchmarks. This enabled precise evaluation of the overall system. As shown in Fig. 5(b), the optimization reduces power and latency by ~80% and ~67% compared to no-skip and traditional zero-value bit skip, respectively.

## IV. Evaluation

As shown in Fig. 6, the layout of the digital macro designed for energy-efficient attention computation is presented. The macro was implemented in a 65-nm process and occupied a total area of 0.35 mm². It had a maximum memory capacity of 64×64×8 bits for the weights. With a 1 V supply voltage, the macro's power consumption is 1.24 mW. At a frequency of 100 MHz, the macro achieved a peak performance of 42.27 GOPS, a maximum energy efficiency of 34.1 TOPS/W, and an area efficiency of 120.77 GOPS/mm².

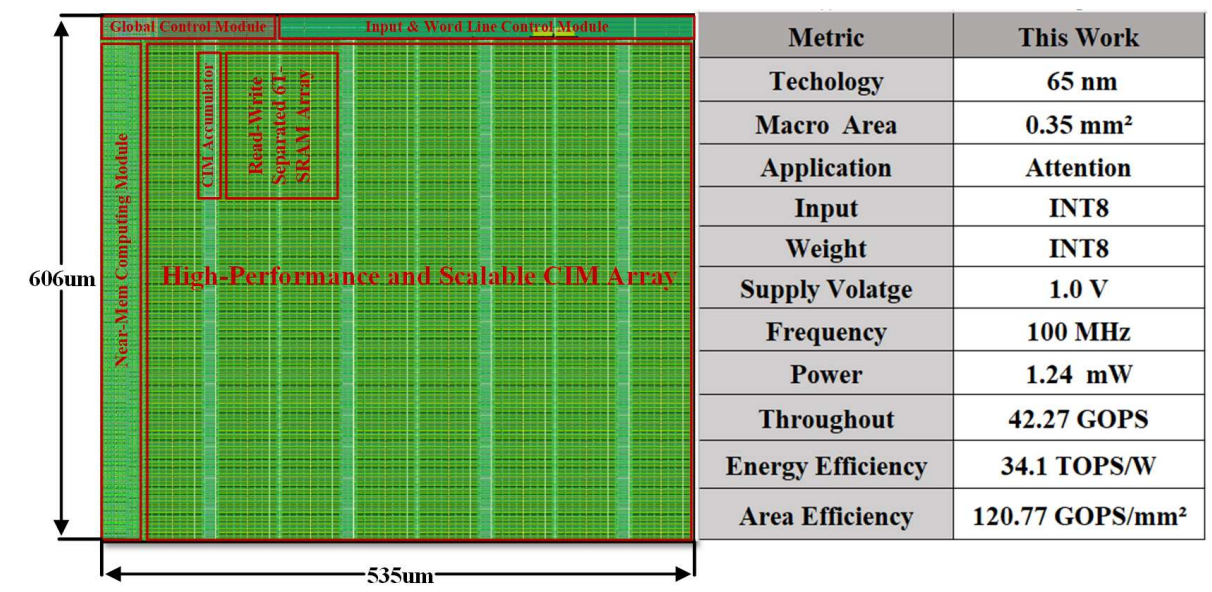

| Metric | This Work |
| --- | --- |
| Techology | 65 nm |
| Macro Area | 0.35 mm² |
| Application | Attention |
| Input | INT8 |
| Weight | INT8 |
| Supply Volatge | 1.0 V |
| Frequency | 100 MHz |
| Power | 1.24 mW |
| Throughout | 42.27 GOPS |
| Energy Efficiency | 34.1 TOPS/W |
| Area Efficiency | 120.77 GOPS/mm² |

Fig. 6. The macro physical layout and specifications.

### *a) Practical Applications and Performance Comparison*

Our experiment involved computing attention scores with the pre-trained ViT and DETR models. The resulting energy consumption data from the process is shown in Fig. 7. We performed these measurements on a CPU (Intel64 Family 6 Model 183 Stepping 1), a GPU (NVIDIA GeForce RTX 4070), and our CIM macro. The results confirmed our CIM macro's superior energy efficiency. In image recognition, it was 25.2× and 12.9× more efficient than the CPU and GPU, respectively. For visual semantic segmentation, this advantage grew to 26.8× over the CPU and 13.3× over the GPU.

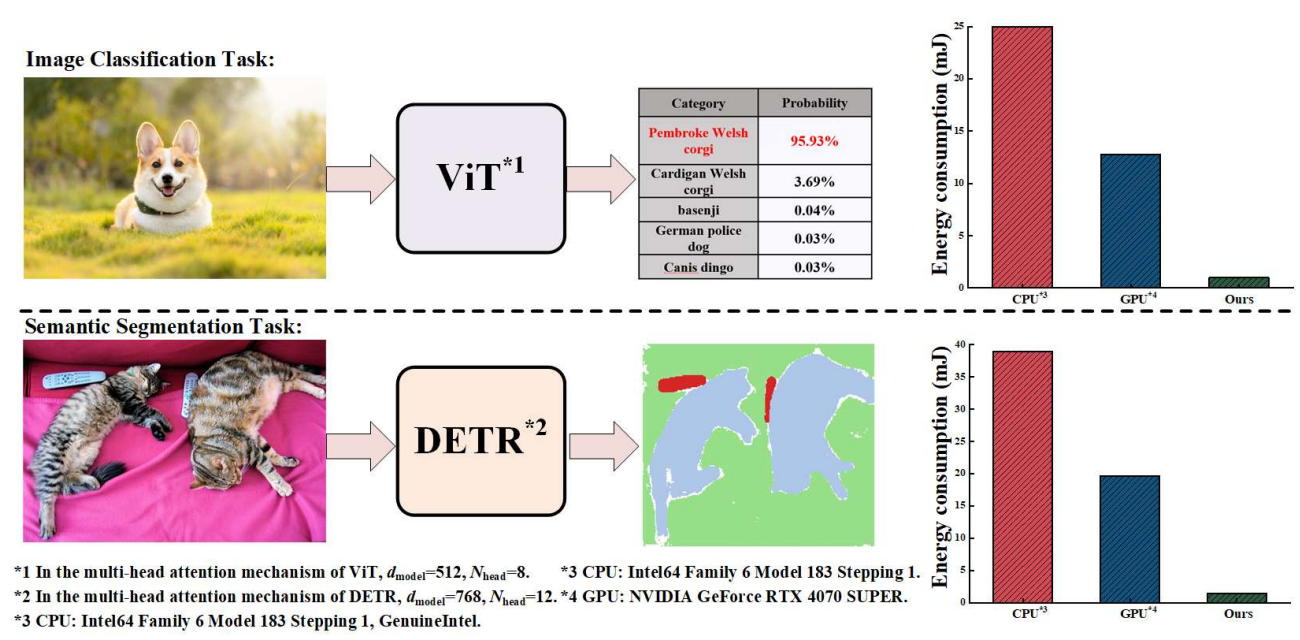

Fig. 7. The comparison of the energy efficiency advantage of our work against CPU and GPU for attention score computation.

Furthermore, we evaluated several Transformer-CIMs in terms of memory accesses and energy consumption, with the results shown in Fig. 8. The evaluation looked at the least memory accesses needed to compute the attention scores and the lowest energy consumption. The baseline for our comparison was traditional CIM accelerators configured for parallel computation. Our experimental results demonstrated that our approach reduces memory accesses by 6.9× and energy

TABLE I. Comparison With The State-of-The Works

| Work | Y. Wang [12] [ISSCC'22] | F. Tu [6] [ISSCC'22] | X. Fu [7] [TCAS-i'23] | S. Liu [14] [ISSCC'23] | R. GUO [9] [JSSC'25] | This work | This work (scaled) |
| --- | --- | --- | --- | --- | --- | --- | --- |
| Architecture | No-CIM | Digital CIM | Digital CIM | Digital CIM | Analog CIM | Digital CIM | Digital CIM |
| Technology (nm) | 28 | 28 | 28 | 28 | 28 | 65 | 28 |
| Macro Area (mm²) | 6.82 | 6.83 | 2 | 3.93 | 7.09 | 0.35 | 0.064*4 |
| Supply Voltage (V) | 0.56 - 1.1 | 0.6 - 1.0 | 0.6 - 1 | 0.64 - 1.03 | 0.56 - 0.9 | 1.0 | 0.8 |
| Frequency (MHz) | 50 - 510 | 80 - 240 | 50 - 200 | 20 - 320 | 80 - 275 | 100 | 100 |
| Power (mW) | 12.06 - 272.8 | 27.04 - 118.2 | 20.42-81.83 | 8.72 - 250.65 | 43.07 - 57.97 | 1.24 | 0.26*3 |
| Peak Performance*1,2 (TOPS) | 4.07 | 0.74 | 0.8 | 3.33 | 1.38 | 0.042 | 0.042 |
| Energy Efficiency*1,2 (TOPS/W) | 27.56 | 20.5 | 23.24 | 25.22 | 19.38 | 34.09 | 161.5*3 |
| Area Efficiency*1,2 (GOPS/mm²) | 596.8 | 108.3 | 400 | 847.3 | 194.4 | 120.77 | 656.25*4 |

*1 Off-chip memory access is excluded. All on-chip computing time consumption is included. *2 One operation indicates one addition or multiplication.
*3 Power scaling from the 65nm to the 28 nm process node is approximately: $P_{28nm} = P_{65nm} \times (28nm / 65nm) \times (V_{28nm} / V_{65nm})^2 \times (f_{28nm} / f_{65nm})$ [15].
*4 Area scaling from the 65nm to the 28 nm process node is approximately: $S_{28nm} = S_{65nm} \times (28nm / 65nm)^2$ [15].

consumption by 4.9× compared to this baseline. Moreover, our CIM macro outperformed other advanced Transformer-CIMs.

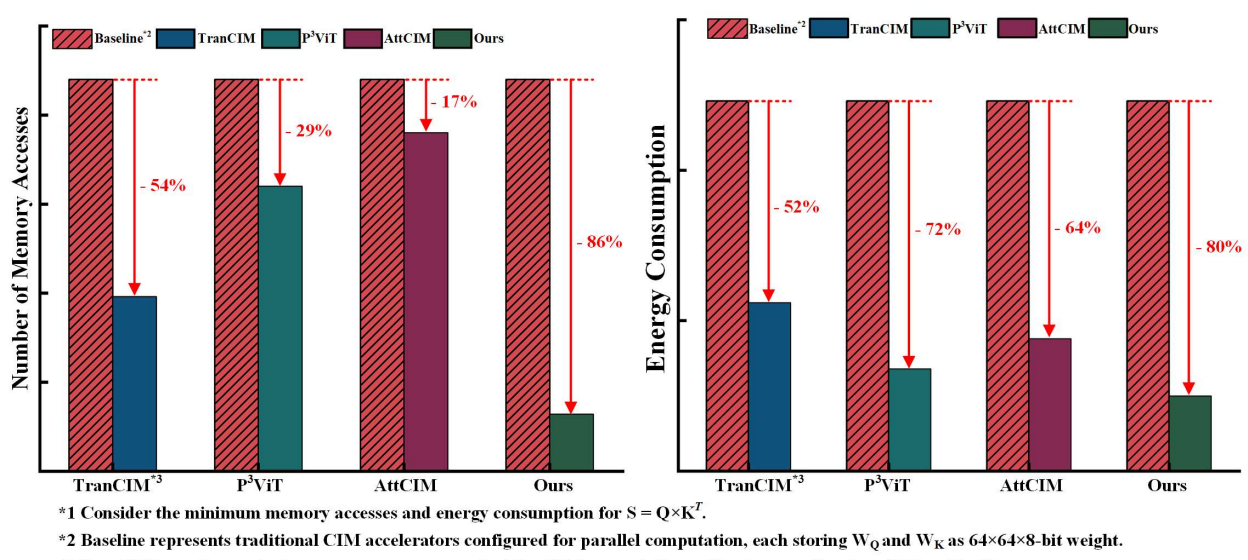

Fig. 8. The performance comparison of various Transform-CIM in terms of memory accesses and energy consumption.

### *b) Comparison With State-of-the-Art Works*

Table I shows a detailed comparison with the state of-the-art Transformer accelerators. The work in [12] is a digital Transformer accelerator that uses approximate computing techniques for attention. TranCIM [6] and P3ViT [7] are digital CIMs with pipelined/parallel reconfigurable mode. A digital CIM with unstructured weight pruning and local attention is proposed in [14]. AttCIM [9] is a digital-assisted CIM array with a CIM ring architecture and data flow reshaping. These works fully leverage the advantages of pipelined structures and the sparsity of data in attention computations to improve energy efficiency.

To ensure a fair comparison, we scaled our work to the 28-nm process node to benchmark against other works. Utilizing the advantages of CIM, our work achieves a 6× improvement in energy efficiency compared to [12]. By mitigating the overhead of dynamic matrix multiplication, our work is at least 7× more energy-efficient than [6], [7], [9], and [14]. Moreover, our design demonstrates a significant area advantage, with an area efficiency at least 2× higher than that of [6], [7], and [9].

## V. CONCLUSION

This work proposes a high-efficiency digital SRAM-based CIM macro for Transformer attention mechanisms, aiming to address the bottleneck of weight-stationary architectures in handling dynamic MM. The attention score computation is reformulated through a combined QK-weight matrix, transforming dynamic MM into static MM to completely eliminate the overhead of frequent memory updates. Simultaneously, to address the resulting 2-input computation challenge, we employ a decomposition scheme of logic, shifting, and addition, combined with a hierarchical zero-value bit skip. This approach deeply exploits data sparsity while avoiding costly physical multipliers, significantly reducing redundant computations. Experimental results show that at the same technology node, our design achieves at least a 7× energy efficiency and 2× area efficiency gain compared to existing state-of-the-art CIM accelerators. These findings fully validate the significant advantages and application potential of this scheme in edge intelligence hardware.

## REFERENCES

[1] Y.-D. Chih et al., "16.4 an 89 TOPS/W and 16.3 TOPS/mm$^2$ all-digital SRAM-based full-precision compute-in-memory macro in 22 nm for machine-learning edge applications," in *IEEE Int. Solid-State Circuits Conf. (ISSCC) Dig. Tech. Papers,* pp. 252–254, Feb. 2021.

[2] X. Si et al., "A local computing cell and 6T SRAM-based computing-in-memory macro with 8-b MAC operation for edge AI chips," *IEEE J. Solid-State Circuits*, vol. 56, no. 9, pp. 2817–2831, Sep. 2021.

[3] F. Tu et al., "A 28nm 29.2TFLOPS/W BF16 and 36.5TOPS/W INT8 Reconfigurable Digital CIM Processor with Unified FP/INT Pipeline and Bitwise In-Memory Booth Multiplication for Cloud Deep Learning Acceleration," in *IEEE Int. Solid-State Circuits Conf. (ISSCC) Dig. Tech. Papers*, pp. 1-3, Feb. 2022.

[4] F. Tu et al., "16.1 MuITCIM: A 28nm 2.24μJ/Token Attention-Token Bit Hybrid Sparse Digital CIM-Based Accelerator for Multimodal Transformers," in *IEEE Int. Solid-State Circuits Conf. (ISSCC) Dig. Tech. Papers*, pp. 248–250, Feb. 2023.

[5] S. Liu et al., "16.2 A 28nm 53.8TOPS/W 8b Sparse Transformer Accelerator with In-Memory Butterfly Zero Skipper for Unstructured-Pruned NN and CIM-Based Local-Attention-Reusable En gine," in *IEEE Int. Solid-State Circuits Conf. (ISSCC) Dig. Tech. Papers*, pp. 250-252, Feb. 2022.

[6] F. Tu et al., "A 28 nm 15.59 μJ/token full-digital bitline-transpose CIM-based sparse transformer accelerator with pipeline/parallel reconfigurable modes," in *IEEE Int. Solid-State Circuits Conf. (ISSCC) Dig. Tech. Papers,* pp. 466–468 , Feb. 2022.

[7] X. Fu et al., "P$^3$ViT: A CIM-based high-utilization architecture with dynamic pruning and two-way ping-pong macro for vision transformer," *IEEE Trans. Circuits Syst. I, Reg. Papers*, vol. 70, no. 12, pp. 4938–4948, Dec. 2023.

[8] R. Guo et al., "CIMFormer: A systolic CIM-array-based transformer accelerator with token-pruning-aware attention reformulating and principal possibility gathering," *IEEE J. Solid-State Circuits*, vol. 59, no. 10, pp. 3317–3329, Oct. 2024.

[9] R. Guo et al., "A 28-nm 28.8-TOPS/W Attention-Based NN Processor With Correlative CIM Ring Architecture and Dataflow Reshaped Digital-Assisted CIM Array, " *IEEE J. Solid-State Circuits*, vol. 60, no. 1, pp. 332-346, Jan. 2025.

[10] S. Qin et *al.*, "StreamDCIM: A Tile-based Streaming Digital CIM Accelerator with Mixed-stationary Cross-forwarding Dataflow for Multimodal Transformer," *IEEE International Symposium on Circuits and Systems (ISCAS)*, pp. 1-5, London, Feb. 2025.

[11] A. Vaswani et al., "Attention is all you need," in *Proc. Adv. Neural Inf. Process. Syst.*, pp. 5998–6008, Nov. 2017.

[12] Y. Wang et al., "A 28 nm 27.5 TOPS/W approximate-computing-based transformer processor with asymptotic sparsity speculating and out-of order computing," in *IEEE Int. Solid-State Circuits Conf. (ISSCC) Dig. Tech. Papers*, pp. 1–3, Feb. 2022.

[13] F. Tu et al., "16.1 MuITCIM: A 28 nm 2.24μJ/token attention-token-bit hybrid sparse digital CIM-based accelerator for multimodal transform ers," in *IEEE Int. Solid-State Circuits Conf. (ISSCC) Dig. Tech. Papers*, pp. 248–250, Feb. 2023.

[14] S. Liu et al., "16.2 A 28 nm 53.8 TOPS/W 8b sparse transformer accelerator with in-memory butterfly zero skipper for unstructured pruned NN and CIM-based local-attention-reusable engine," in *IEEE Int. Solid-State Circuits Conf. (ISSCC) Dig. Tech. Papers*, pp. 250–252, Feb. 2023.

[15] Stillmaker et al., "Scaling Equations for the Accurate Prediction of CMOS Device Performance from 180nm to 7nm." *Integration the VLSI Journal.* vol. 5, pp. 74-81, Jan. 2017.